\documentclass[aps,prl,twocolumn]{revtex4-1}

\usepackage{dcolumn}
\usepackage{amsmath}
\usepackage{amsthm}
\usepackage{amstext}
\usepackage{orcidlink}
\usepackage{graphicx}

\def \Tr {\mathop{\rm Tr}\nolimits}
\def \PT {\mathop{\rm PT}\nolimits}
\def\T{T} 
\def\nn{\notag} 

\usepackage{color}

\begin{document}
\title{Two-Loop Spacelike Splitting Amplitude for  $\mathcal N=4$ Super-Yang-Mills Theory}



\author{
 Johannes Henn,$^{a}$ Rourou Ma$^{d}$, Yongqun Xu$^{d}$, Kai Yan,$^{f,g}$ Yang Zhang\orcidlink{0000-0001-9151-8486},$^{d,e}$ Hua Xing Zhu$^{h,i}$
}

\affiliation{
$^a$ Max-Planck-Institut f\"ur Physik, Werner-Heisenberg-Institut, 85748 Garching bei M\"unchen, Germany \\
$^d$ Interdisciplinary Center for Theoretical Study, University of Science and Technology of China, Hefei, Anhui 230026, China \\
$^e$ Peng Huanwu Center for Fundamental Theory, Hefei, Anhui 230026, China\\
$^f$ School of Physics and Astronomy, Shanghai Jiao Tong University, Shanghai 200240, China \\
$^g$ Key Laboratory for Particle Astrophysics and Cosmology (MOE), Shanghai 200240, China \\
$^h$ School of Physics, Peking University, Beijing, 100871, China \\
$^i$ Center for High Energy Physics, Peking University, Beijing 100871, China
}

\date{\today}
\begin{abstract}
The study of collinear behavior for gauge theories in the spacelike region is of great phenomenological and theoretical importance. We analytically calculate the two-loop spacelike splitting amplitude for the full color $\mathcal N=4$ Super-Yang-Mills theory. 
The result is derived by two complementary methods starting from the known amplitude: one is based on a discontinuity analysis, while the other one is based on calculating Feynman integrals analytically in the collinear limit.
%
Our result explicitly shows terms that violate naive factorization.
However we show that 
factorization is restored at the level of color-summed unpolarized squared amplitudes at next-to-next-to-next-to leading order.  We conjecture that the two-loop tripole terms in the generalized splitting amplitudes in QCD are identical to what we obtain in $\mathcal{N}=4$ super Yang-Mills theory.

\end{abstract}

\maketitle


\section{Introduction}

The factorization of perturbative quark-gluon dynamics from non-perturbative hadron dynamics lies at the heart of high-energy collider physics, such as the Large Hadron Collider (LHC). It allows the prediction of hard scattering cross sections, denoted by $\sigma$, through the schematic formula:
\begin{equation}
\label{eq:sigma_fact}
\sigma = f \otimes f \otimes \hat{\sigma} \,,
\end{equation}
where $f$ represents the Parton Distribution Functions (PDFs) parameterizing the non-perturbative partonic structure of hadrons, and $\hat{\sigma}$ is the partonic cross section calculated from on-shell scattering amplitudes of quarks, gluons, and other Standard Model particles. Building on the factorization theorem, impressive results have been achieved for the precision program at the LHC, examples ranging from Next-to-Next-to-Leading Order~(NNLO) high multiplicity processes~\cite{Czakon:2021mjy,Agarwal:2021grm,Chawdhry:2021hkp,Kallweit:2020gcp,Badger:2023mgf,Badger:2021ohm} to Next-to-Next-to-Next-to-Leading Order~(NNNLO) color singlet production at the LHC~\cite{Anastasiou:2015vya,Mistlberger:2018etf,Dulat:2018bfe,Cieri:2018oms,Dreyer:2018qbw,Duhr:2019kwi,Chen:2019lzz,Duhr:2020sdp,Duhr:2020seh,Chen:2021vtu,Billis:2021ecs,Chen:2021isd,Camarda:2021ict,Chen:2022cgv,Chen:2022lwc}.

The predictive power of the factorization theorem in Eq.~\eqref{eq:sigma_fact} relies on the universality of the PDFs, meaning that they should be independent of the underlying scattering processes. Given the importance of factorization, it has been studied with great efforts since the early days of QCD, resulting in a remarkable proof of factorization for unpolarized Drell-Yan production at hadron colliders~\cite{Collins:1985ue,Collins:1989gx}. For other important processes such as jet production at hadron colliders, a rigorous proof of factorization is currently absent.

In recent years, the study of high-energy scattering in extreme kinematics, notably scattering in the transverse-momentum-dependent region, has revealed an interesting possibility of factorization violation~\cite{Bacchetta:2005rm,Bomhof:2004aw,Bomhof:2006dp}. In particular, an explicit counter-example has been found starting from one-loop in a toy model for single-spin asymmetry in dihadron production at hadron colliders by Collins and Qiu~\cite{Collins:2007nk}. They also suggest that a class of active-spectator diagrams at two loops can potentially lead to cross-section-level violation of collinear factorization for unpolarized dihadron production at N$^3$LO. Related but independently, Catani, de Florian, and Rodrigo show that infrared poles of spacelike splitting amplitudes for multi-jet production can depend on the color and kinematic information of non-collinear partons at two loops~\cite{Catani:2011st}, manifestly violating amplitude-level collinear factorization. They argue that after integrating over the collinear phase space, such non-factorization contributions can lead to process-dependent collinear singularities that cannot be canceled by PDF renormalization, thereby potentially invalidating the universality of PDFs for such processes at sufficiently high order.

In both cases, the origin of potential factorization breaking comes from the loop corrections to the spacelike collinear limit, where a collinear parton is emitted from the incoming parton. Using soft-collinear effective theory for forward scattering~\cite{Rothstein:2016bsq}, such effects have been related to the exchange of Glauber gluons~\cite{Schwartz:2017nmr}. Explicitly, for a massless scattering process $p_1 + p_2 \to p_3 + p_4 + p_5$, the amplitude in the spacelike collinear limit, $p_2 \parallel p_3$, factorizes as:
\begin{equation}
\label{eq:spacelike}
\mathcal{A}_5 (p_1, p_2, p_3, p_4, p_5) \xrightarrow{p_2 \parallel p_3 } \mathbf{Sp} \times
\mathcal{A}_4
(p_1, \tilde{P} ,p_4, p_5) \,.
\end{equation}
In the usual strict collinear factorization, the splitting amplitude $\mathbf{Sp}$ depends on the quantum numbers and kinematics of the collinear pair only. However, Ref.~\cite{Catani:2011st} found that for spacelike collinear limits such as those in \eqref{eq:spacelike}, the splitting amplitude necessarily involves the quantum numbers and momenta of non-collinear partons.

Given the importance of the collinear factorization for collider physics, significant effort has been spent examining the consequences of the generalized collinear factorization formula in \eqref{eq:spacelike}. Interestingly, the pole terms of the amplitude-level factorization violation found in \cite{Catani:2011st} cancel at the squared cross-section level, owing to a form of exponentiation of the infrared singularities. For non-pole terms, the soft limit of the generalized splitting amplitude has been studied in \cite{Dixon:2019lnw}, where a cancellation at the cross-section level has also been found.
Recently, amplitude-level factorization violation has also been investigated in the context of more than one collinear direction \cite{Cieri:2024ytf}.

\begin{figure}[t!]
    \centering
    \includegraphics[width=\linewidth]{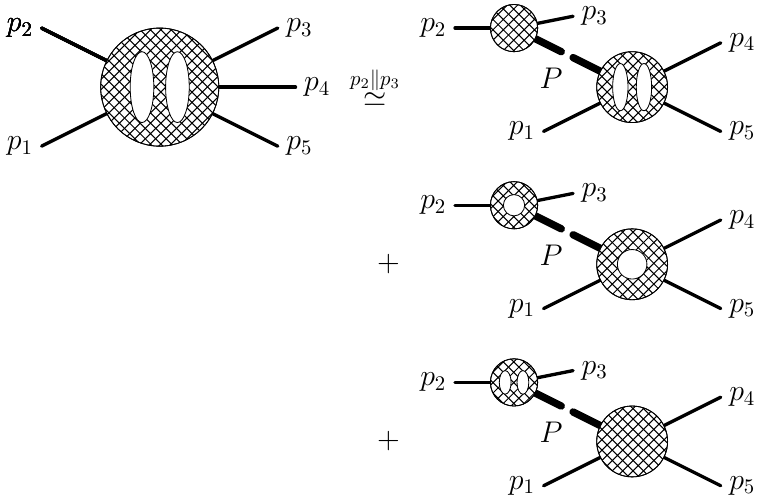}
    \caption{Schematic form of spacelike collinear factorization for two-loop five-particle scattering. The full factorized expression involves the generalized spacelike splitting amplitude at tree, one- and two loops, as well as the $2\to 2$ amplitudes from tree to two loops.}
    \label{fig:split}
\end{figure}

Despite these efforts, a conclusive statement about the possibility of factorization breaking or not at the cross-section level from spacelike splitting such as \eqref{eq:spacelike} is still missing. In this \emph{Letter}, we initiate a systematic study in this direction by computing, for the first time, the \emph{finite} terms in the generalized splitting amplitude in \eqref{eq:spacelike} to two loops, as depicted in Fig.~\ref{fig:split}. 
We perform the calculation in ${\cal N} = 4$ Super Yang-Mills Theory (${\cal N} = 4$ SYM), a close cousin of QCD that shares similar infrared behavior in perturbative theory. Building upon the remarkable data for high multiplicities amplitudes in the literature~\cite{Chicherin:2018old,Chicherin:2020oor,Henn:2021cyv,Henn:2024ngj,Agarwal:2023suw,DeLaurentis:2023nss}, we systematically develop techniques for the analytic continuation of high-multiplicity amplitudes from timelike collinear kinematics to spacelike collinear kinematics. We believe such techniques will be useful in future investigations of more complicated factorization-breaking configurations.

While the intermediate steps of our calculation involve lengthy expressions, remarkably, the final generalized splitting amplitudes in ${\cal N} = 4$ SYM have a simple form. More importantly, the potential factorization-breaking terms in the generalized splitting amplitude enjoy a partially exponentiated form, causing the factorization-breaking effects to cancel at the cross-section level.

This \emph{Letter} is organized as follows. We first calculate the two-loop five-point massless integrals in the spacelike collinear limit in terms of Goncharov polylogarithms (GPLs) \cite{GoncharovGPL} and then assemble the $\mathcal N=4$ two-loop five-point amplitude in this limit. These computations provide a solid ground for the colliner factorization violating discussion. In parallel, we analyze the analytic continuation of five-point scattering amplitudes, and explicitly identify the collinear factorization violating terms. This computation agrees with the result from our master integral computations. Based on 
these
computations, we finally derive the generalized 
two-loop  
splitting amplitudes, which constitutes our main result. 

\section{Spacelike collinear limit of two-loop five-point massless Feynman integrals}
 In this section, we calculate all  two-loop five-point massless master integrals in the spacelike collinear region, up to weight $4$ in terms of GPLs. The result will be  used for calculating the $\mathcal N=4$ two-loop five-point amplitude and then the two-loop splitting function in this limit. 

The $2\to 3$ scattering  for massless particles is characterized by the five Mandelstam variables $s_{12}$, $s_{23}$, $s_{34}$, $s_{45}$ and $s_{15}$. The physical region is,
\begin{equation}
    s_{12}>0, \quad  s_{23}<0, \quad s_{34}>0, \quad s_{45}>0, \quad s_{15}<0, \quad 
    \label{eq:scattering region}
\end{equation}
and $\text{Im}(\epsilon_5)>0$, where $
     \epsilon_{5}\equiv 4 i \ \varepsilon_{\mu_1 \mu_2 \mu_3
    \mu_4} p_1^{\mu_1} p_2^{\mu_2}p_3^{\mu_3}p_4^{\mu_4}$.

There are three types of two-loop five-point massless Feynman integral families, namely pentagon-box, hexagon-box and double pentagon. The corresponding canonical differential equation was obtained in ref.~\cite{Gehrmann:2018yef, Chicherin:2018mue, Abreu:2018aqd, Chicherin:2018old}. There are $31$  letters for the integrals~\cite{Chicherin:2017dob}. The boundary values and the iterative integral form were calculated in \cite{Chicherin:2018old}. The analytic boundary value at a point
\begin{equation}
X_0:\quad \{s_{12},s_{23},s_{34},s_{45},s_{15}\}=\{3, -1, 1, 1, -1\}
\end{equation}
is available in ref.~\cite{Chicherin:2020oor}. 
Furthermore in the $12\to 345$ scattering region, up to weight 2, all master integrals are obtained in terms of classical polylogarithms, while the weight $3$ and $4$ parts are expressed as one-fold integrals for the fast numeric evaluation~\cite{Chicherin:2020oor}.

Our goal is to get  all two-loop five-point massless integrals up to the weight $4$ in terms of polylogarithms, in the spacelike collinear region. Without loss of generality, we consider $2 \parallel 3$, and a generic point $P$ in this region can be parameterized as,
\begin{equation}
\{s_{12},s_{23},s_{34},s_{45},s_{15}\}=\{s z, -4 \delta^2, (1-z) x s, s, x s + c \delta\}
\label{eq:collinear_region}
\end{equation}
with the parameter range $s>0$, $z>1$, $x<0$, $\delta\to 0$.

\begin{figure}
    \centering
\includegraphics[width=\linewidth]{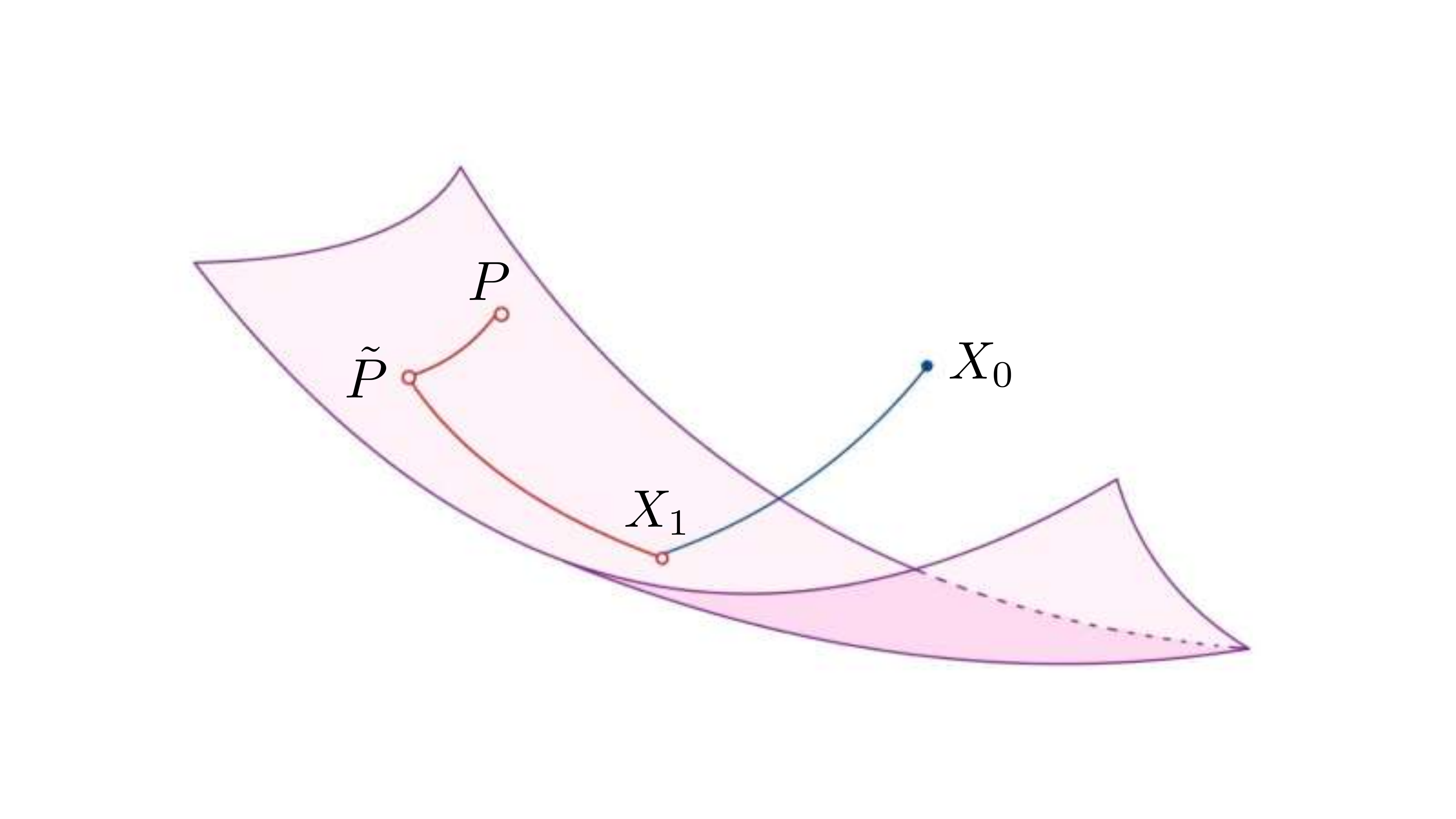}
    \caption{Path for solving the canonical differential equation of two-loop five-point integrals in the spacelike region. One starts from the base point $X_0$, and integrates along a contour to a special point $X_1$ on the $2||3$ collinear region. Two further integrations are taken to reach the generic collinear region point $P$.}
\label{fig:2l5p_int}
\end{figure}

We solve the canonical differential equations along the integration path illustrated in Fig.\ref{fig:2l5p_int}.
It is important to get the boundary value at a point in that region. This point is chosen as,
\begin{equation}
    X_1:\quad  \{s_{12},s_{23},s_{34},s_{45},s_{15}\}= \{4,-4\delta^2,1,2,-1\}
\end{equation}
where $\delta$ is a small positive number. Boundary value at $X_1$ can be determined by solving the canonical differential equation, along a curve from $X_0$ to $X_1$, 
\begin{equation}
    \{s_{12},s_{23},s_{34},s_{45},s_{15}\}=\{\frac{4}{\lambda^2+1}, \frac{-4\lambda^2}{\lambda^2+1}, 1, \frac{2-2\lambda^2}{\lambda^2+1}, -1\}
\end{equation}
The base point $X_0$ corresponds to $\lambda=1/\sqrt 3$, while $X_1$ corresponds to $\lambda\to \delta$. It is straightforward to solve the differential equation along this curve to obtained 
boundary values at $X_1$, which are polynomials of $\log\delta$. 

A high-precision PSLQ computation determines the coefficients of $\log\delta$ for all boundary values at $X_1$ as the combinations of   constants used in ref.~\cite{Chicherin:2020oor} for the boundary value at $X_0$. Therefore the boundary values at $X_1$ are simplified to a compact form.
With  analytic boundary values, 
consider two steps,
\begin{itemize}
    \item Step 1. Integrate from $X_1$ to a point $\tilde P$ in the spacelike collinear region with $c=0$,
    \begin{equation}
\{s_{12},s_{23},s_{34},s_{45},s_{15}\}=\{s z, -4 \delta^2, (1-z) x s, s, x s \}
    \end{equation}
Fix $s$ as a constant, and  the $31$ letters are reduced to $10$  rational letters in the limit $\delta\to 0$. 
\begin{gather}
x, 1 + x, -1 + z, z, -1 - x + z, x + z, 
 1 + x z, \nonumber \\
 -1 + z + x z, -1 - x + x z, -x + z + x z
\end{gather}
Then we solve the differential equation along a path from $X_1$ to $\tilde P$, and integrate $x$ and $z$. The dependence on $s$ is restored via dimensional analysis.

    
\item Step 2. Then we carry out the integration for the variable $c$ to get the master integrals evaluated at a generic point $P$ in the spacelike collinear region. The 
following substitution allows us to rationalize $\epsilon_5$:
\begin{equation}
\quad \quad c \equiv \frac{-8 s^2 x (1 + x) y z\sqrt{-s^3 x (1 + x)(-1 + z)z}}{1 + y^2},
\label{eq:eulersubstitute}
\end{equation}
Here the new variable $y$ is related to the cross ratios $z_I$ and $\bar{z}_I$ from \eqref{eq:z definition} as follows, 
\begin{equation}
\left(\frac{i+y}{i-y}\right)^4=\frac{z_4 z_5}{\bar{z}_4 \bar{z}_5}
\end{equation} 
The physical scattering condition in \eqref{eq:scattering region} implies that $y$ is real and small. 
Next, we integrate over $y$ along the real axis,
where
the $31$ letters are reduced to four rational letters,
    \begin{equation}
        {1 + y, -1 + y, i + y, -i + y}
    \end{equation}
The integration in $y$ along the real axis is thus straightforward. We note that in the limit $\delta\to 0$, only a subset of nonplanar Feynman integrals depend on $y$, while all planar integrals do not.
\end{itemize}
Finally we get all two-loop five-point master integrals, in terms of GPLs, up to weight $4$, for generic points in the spacelike collinear region. This result is verified with the numerics from package {\sc PentagonMI} \cite{Chicherin:2020oor}, in the limit $\delta \to 0$.




\section{Two-loop five-point $\mathcal N=4$ scattering amplitude in the spacelike collinear region}

The full-color two-loop five-point $\mathcal N=4$ scattering amplitude can be constructed from the  integrand in ref.~\cite{Carrasco:2011mn}. The corresponding color basis has $12$ single trace  and $10$ double trace functions~\cite{Edison:2011ta},
\begin{eqnarray}
    \T_1 	&=&  \left[\Tr (12345) - \Tr(15432)\right], \quad \ldots \nonumber \\
\T_{13} &=&  \Tr (12) \left[ \Tr(345) - \Tr(543) \right],  \quad \ldots
\label{eqn:5pt_color_basis}
\end{eqnarray}
The MHV amplitude can be expanded over the $22$ color functions, and $6$ Parke-Taylor factors,
\begin{gather}
    \PT_1 = \PT(12345), \quad \PT_2 = \PT(12354),\nonumber \\
\PT_3=\PT(12453),\quad \PT_4=\PT(12534),\nonumber \\
\PT_5=\PT(13425),\quad \PT_6=\PT(15423)
\end{gather}
with the definition,
\begin{equation}
\PT(i_1 i_2 i_3 i_4 i_5)=\frac{\delta^8(Q)}{\langle i_1 i_2 \rangle\langle i_2 i_3 \rangle\langle i_3 i_4 \rangle\langle i_4 i_5 \rangle\langle i_5 i_1 \rangle}
\end{equation}
where $\delta^8(Q)$ is the Dirac delta function for the superspace.
The two-loop five-point $\mathcal N=4$ amplitude's symbol expression was calculated in \cite{Abreu:2018aqd, Chicherin:2018yne}. On the other hand, the analytic two-loop (and three-loop) four-point $\mathcal N=4$ super-Yang-Mills amplitude is  in ref.~\cite{Henn:2016jdu}.

With our analytic result of two-loop five-point master integrals in the spacelike collinear limit $2\parallel 3$, it is straightforward to  
assemble the $\mathcal N=4$ amplitude in this region. We also consider the ratio between Parke-Taylor factors in this limit,
\begin{equation}
    \left\{\frac{\PT_1}{\PT_1},\ldots,\frac{\PT_6}{\PT_1}\right\}\underset{\delta\to 0}{\sim}
    \ \left\{1,\frac{-x}{x+1},0,0,0,1\right\}
\end{equation}
with the parametrization defined in \eqref{eq:collinear_region}.

The analytic result for $\mathcal N=4$ amplitude in the spacelike collinear limit, consists of only classical polylogarithms. We achieved the simplification by firstly rewriting the amplitude in terms of harmonic polylogarithms (HPLs) and then decomposing them into Lyndon words with the aid of the package {\sc Polylogtools} \cite{Duhr:2019tlz}. The dependence on the parameter $y$ is only through the power of the function $\log\left(\frac{i- y}{i+ y}\right)$.

The amplitude's analytic expression is provided in the supplemental material.


\section{Discontinuity of the five-point $\mathcal N=4$ scattering amplitude}

\begin{figure}
\centering
\includegraphics[width=\linewidth]{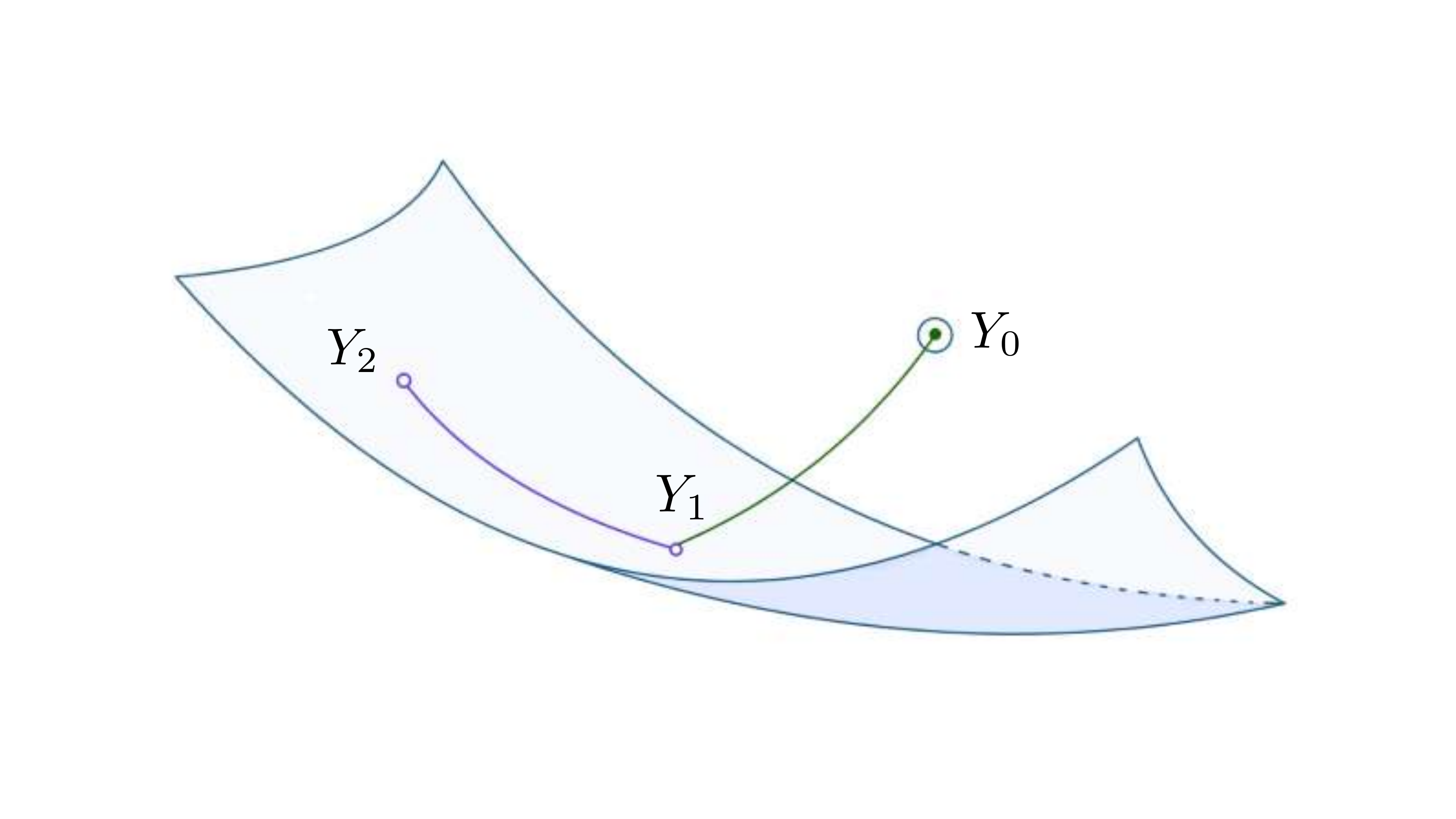}
    \caption{Path for analytic continuation of the five-point $\mathcal N=4$ amplitude. We start from a generic point $Y_0 $ with $\tau>0$ and pick up the monodromy around $\tau=0$ before taking the $2 \parallel 3$ collinear limit.  Then we travel from a timelike region point $Y_1 \, (\tau>0)$ to the spacelike region point $Y_2 \, (\tau<0)$.  }
     \label{fig:discontinuity}
\end{figure}

Gauge theory amplitudes factorize in a universal way in the timelike collinear regime \cite{Bern:1994zx,Bern:1995ix,Kosower:1999xi,Feige:2014wja}. 
In this section we describe a procedure to obtain the spacelike collinear limit of the five-point amplitudes from the time-like regime via analytic continuation.
This approach is complementary to the method we exploit in previous sections. 

Assuming timelike splitting, 
the (color-stripped)  
 five-point amplitude strictly factorizes \cite{Bern:1998sc,Bern:1999ry,Bern:2004cz,Badger:2004uk}
\begin{align}
A_5|_{\text {T.L.}} (2, 3,\cdots ) \xrightarrow{2 \parallel 3}  \text{Split}_{-\lambda} (z; 2, 3)\, A_4 (P^\lambda, \cdots  )
\end{align}
In the collinear kinematic space, the amplitude is governed by a universal  splitting function depending on  the momentum fraction variable $z$ ranging from 0 to 1. 
Analytic continuation of the splitting function into the spacelike regime $(z>1)$ is obstructed by the ambiguity in the sign of its imaginary part \cite{Bern:2004cz}. 

This ambiguity can be traced back to the sign of $i \pi$'s generated in different discontinuity channels, e.g.
\begin{align}
\ln \frac{s_{3I}}{s_{PI}} \cong   \ln |1-z| + i \pi \,\theta(z-1)\, \text{sign}(s_{PI})
\end{align}
which relies on the signature of non-collinear leg $I$ in the scattering process.
To resolve this issue we propose a path for analytic continuation for the full amplitudes, which  
starts from a generic point $Y_0$ away from the collinear kinematic space where $\delta=0$.
The path is parameterized by a variable 
\begin{align}
\tau \equiv s_{13 }/(s_{1 3}+ s_{12} )
\end{align}
where subscript '$1$' labels the unique incoming non-collinear leg in the five-point scattering process. Note that  a generic kinematic point can be parameterized by $\{\tau, s, \delta, c, x \}$ whereas in the collinear limit $\tau$ reduces to $1-z$. 
Starting from a point on the positive real $\tau-$axis, 
the analytic continuation path $\Gamma$ contains three segments: 
\begin{align}
\Gamma_1 :   \tau \rightarrow \tau e^{-2  \pi i } , \quad 
\Gamma_2:  \delta \rightarrow 0 \,, 
\quad 
\Gamma_3 :  \tau \rightarrow -|\tau| - i 0 \,.  \nn 
\end{align}
$\Gamma_1$ is a  residue contour encircling the origin of the real $\tau-$axis, picking up the monodromy around $\tau=0$. Wrapping the contour along $\Gamma_1$,  the discrepancy between the signature of non-collinear momenta has been compensated.   
In the next step we send $\delta $ to zero, landing onto the a point $Y_1$ in the timelike collinear region. 
In the final step, the contour $\Gamma_3$ lies  within the collinear kinematic space. It goes from the positive to the lower side of negative real $\tau-$axis, reaching a point $Y_2$ in the spacelike collinear region. 
The full path for the analytic continuation is illustrated in Fig.\ref{fig:discontinuity}.

The discontinuity along the path $\Gamma$ accounts for the difference between the five-point amplitudes in the timelike v.s. spacelike collinear regime, which define the factorization breaking terms. Schematically we have 
\begin{align}
A_5|_{\text{S.L.}}  \cong A_5|_\text{T.L.} + \text{disc}_\Gamma [A_5] 
\end{align}
A shortcut to analyzing the discontinuity is 
to work with the symbol of the amplitudes. We start with the symbol of the one- and two-loop five-point amplitudes \cite{Abreu:2018aqd,Chicherin:2018yne},
promoting them to the maximally iterated coproduct by the restoring the $i\pi$'s associated with its first entry \cite{Duhr:2012fh}. 
Then we  perform analytic continuation on the coproduct following the path $\Gamma$ and compare with \cite{Dixon:2019lnw}. This allows us to determine the two-loop factorization breaking terms (see Eq.~\eqref{eqn:splitAmpTwoLoop}) up to potential $\pi^3 \times \mathcal{O}(z-1)$ corrections. 
We find agreement with the explicit computation of two-loop five-point master integrals.

\section{Two-loop generalized splitting amplitudes }

As a result of the analysis in the previous sections, we arrive at the main result of this {\it Letter}. We find that full-color two-loop amplitudes have the following generalized factorization form in the collinear limit where $p_a \cong (1-z) P$,  $p_b \cong z P$,
\begin{align}
\mathcal{A}_5 (p_a, p_b, p_i, p_j, p_k) \xrightarrow{a \parallel b}  \mathbf{Sp} \times 
\mathcal{A}_4 
(P, p_i,p_j, p_k) \,.
\end{align}
 Here $\mathbf{Sp}$ are the generalized splitting amplitudes proposed by Catani et al. \cite{Catani:2011st}. 
We present the one- and two-loop splitting amplitudes 
explicitly through $\mathcal{O}(\epsilon^0)$.
The divergent terms at two loops were already known, but the finite terms are new.

\begin{widetext}

\begin{align}
\mathbf{Sp}^{(1)}  &= \left[ \frac{\mu^2 \,z }{  s_{ab} \, (1-z) } \right]^{\epsilon}  \bigg\{  2 N_c   \, \overline{r}_S^{(1)} (z + i0)    
       +     \mathbf{T}_a \cdot \mathbf{T}_{\text{in}}   \;  ( 2 \pi i )\,    c_1  (\epsilon) \frac1\epsilon  
        \bigg\} \, \mathbf{Sp}^{(0)} \,,
        \label{eqn:splitAmpOneLoop}
\end{align}

\begin{align}
\mathbf{Sp}^{(2)}  &= \left[ \frac{\mu^2\, z}{  s_{ab} \,(1-z)} \right]^{2\epsilon}   \bigg\{   4N_c^2   \,  \overline{r}_S^{(2)} (z + i 0)  \,  \nn  \\
      & + N_c\, \mathbf{T}_a \cdot \mathbf{T}_{\text{in}} \;  ( 2 \pi i )\, \Bigg[   c_2(\epsilon)\,
   \frac{1}{\epsilon^3}  +  c^2_1(\epsilon)\left(  - \frac{2}{\epsilon^2}  \ln z  +  \frac{2}{\epsilon} \ln z \ln \big(\frac{z}{z-1}\big)
 \right.  
 \left.
   -2\,   {\rm Li}_{3} \big( 1- \frac1z \big)   -   \ln (z) \ln^2 \big(\frac{z}{z-1}\big)   
    \right) \Bigg]  \nn \\
     &   +  \sum_{I \in \text{outgoing} } [\mathbf{T}_a \cdot \mathbf{T}_{\rm in} ,   \mathbf{T}_a \cdot \mathbf{T}_{I} ]  \, (2 \pi i)  
     \left[ \left( \frac{1}{2 \epsilon^2} -{ \frac12} \zeta_2 \right) ( \ln   |z_I|^2 + i \pi)  + \frac16  \Big( \ln^2 \frac{z_I}{\bar  z_I}  + 4 \pi^2  \Big) \ln \frac{z_I}{\bar  z_I} +  2 \zeta_3  \right]   \nn \\  
    &      +  \sum_{I \in \text{outgoing} } \{ \mathbf{T}_a \cdot \mathbf{T}_{\rm in} ,   \mathbf{T}_a \cdot \mathbf{T}_{I} \}  \, ( 2\pi^2)   \left[ \frac{1}{2\epsilon^2}  -  {\frac12} \zeta_2  \right]      \bigg\} \,\mathbf{Sp}^{(0)}   \,.
    \label{eqn:splitAmpTwoLoop}
\end{align}

\end{widetext}


We adopt the color space formalism for the color charge operators \cite{Catani:1996vz}.
The subscript 'in' labels an incoming hard particle, and '$I$' labels any of the outgoing hard particle.  
$(z_I, \bar{z}_I)$ are the short-hand notation for the collinear limit of the  cross ratios, 
\begin{align}
z_I  =  \frac{\langle a b \rangle \langle {\rm in}\,  I \rangle  }{  \langle {\rm in}\,  a \rangle  \langle b  I \rangle}, \quad  \bar{z}_I  = \frac{[ a b ] [ {\rm in}\,  I ]  }{  [ {\rm in}\,  a ]  [ b  I ]}\,,   \quad  a\parallel b \,.  
\label{eq:z definition}
\end{align} 
In the physical scattering regime, $z_I$ and $\bar z_I$ are a complex conjugate pair.
In the collinear limit, $z_I, \bar{z}_I \rightarrow 0 $, but the ratio $z_I/\bar{z}_I$ is kept finite. 

The remaining terms in eq. (\ref{eqn:splitAmpTwoLoop}) are best explained by considering, without loss of generality, 
the five-point kinematics specified in eq.~\eqref{eq:scattering region} where $b=2, a=3,\, \text{in}=1, I \in \{ 4,5\}$. 
In that case,
the magnitudes of $\{ z_4, z_5$\}  
are given in terms of the kinematic variables $\{s,  \delta, z,  x \}$ introduced in Eq.~\eqref{eq:collinear_region} as follows
\begin{align}
|z_4|^2 = \frac{4\delta^2\, (1+x)}{s z (1-z) \, x },\quad |z_5|^2 = \frac{4\delta^2\, x}{s z (1-z)\, (1+x)}
\end{align}
The phases of  $\{ z_4, z_5\}$ are parametrized by the variable $y$ introduced in Eq.~\eqref{eq:eulersubstitute}.
Their values differ by a factor of $\pi$, which can be specified in terms of GPLs
\begin{align}
 \frac{1}{2i} \ln \frac{z_4}{\bar{z}_4} 
&=  -i \,G_{-i} (y)+ i \,G_{i} (y)  - \frac{\pi}{2} , \quad \nn\\
\frac{1}{2 i} \ln \frac{z_5}{\bar{z}_5} 
&= -i \,G_{-i} (y) +i \, G_{i} (y)   +  \frac{\pi}{2}
\end{align}

In addition,  the constants $c_L(\epsilon)$'s that appear in Eq.~\eqref{eqn:splitAmpOneLoop} and \eqref{eqn:splitAmpTwoLoop} 
are specified in the following, 
\begin{align}
c_1 (\epsilon) &= -e^{\gamma_E \epsilon} \frac{\Gamma(1+\epsilon)\Gamma^2 (1- \epsilon)}{\Gamma(1- 2 \epsilon)}  \nn \\
c_2 (\epsilon) & =   [c_1(\epsilon)]^2  \frac{ \pi \epsilon }{ \tan (\pi \epsilon) } 
+ \epsilon^2 f (\epsilon) \,  c_1 (2 \epsilon) . 
\end{align}
where $f(\epsilon)= (\psi(1-\epsilon)-\psi(1))/\epsilon $ with $\psi$ the digamma function.

The generalized splitting amplitudes $\mathbf{Sp}$ apply to the case of space-like splitting, where an incoming particle $b$ emits an outgoing collinear particle $a$.  
Setting $\mathbf{T}_a =0$, $\mathbf{Sp}$ reduces to the color-singlet timelike splitting amplitude, which is strictly factorized. Up to two-loop order, the factorization violating effects are associated with a color dipole $\mathbf{T}_a \cdot \mathbf{T}_{\text{in}} $, as well as two types of color tripole built from commutator and anti-commutator between dipole operators. The structure of $\epsilon-$poles in the dipole and tripole functions follows from IR exponentiation, which has been discussed in \cite{Catani:2011st}. Their \emph{finite} part is new.

 Let us discuss this result. In the $L-$loop generalized splitting amplitude,  the color singlet function $\overline{r}_S^{(L)}$ is the universal splitting kernel \cite{Bern:2004cz}
\begin{align}
\overline{r}_S^{(L)} (z) =
 \left[ \frac{s_{ab} \,(1-z)}{ \mu^2 z  } \right]^{L\epsilon}   r_S^{(L), \mathcal{N}=4} (z, s_{ab})
\end{align}
They could be obtained in the case of timelike splitting, then analytically continued to the region where $z$ is greater than one (but carries a small positive imaginary part).

The $1-$ and $2-$loop dipole functions are purely imaginary, and equal to minus $2$ times the imaginary part of the splitting kernel 
$\overline{r}^{(1)}_S (z + i 0) $ and $\overline{r}^{(2)}_S (z+ i 0) $ .

The tripole functions are demonstrated in the last two lines in Eq.~\eqref{eqn:splitAmpTwoLoop}. They correlate the collinear particle $a$ with both an incoming and an outgoing hard particle, which appear for the first time at the level of two-loop five-point amplitudes, originating from the non-planar topologies. 

Intriguingly, their expressions are \emph{identical to} the two-loop tripole soft gluon emission factors in the space-like collinear regime \cite{Dixon:2019lnw}\footnote[1]{A typo was corrected in the updated version of \cite{Dixon:2019lnw}}. 
We observe that the
two-loop tripole functions have no explicit dependence on the $z-$variable which defines the fractions of momenta carried by the collinear pair.
In retrospect, therefore
they could 
have been fixed by comparing to
the soft-collinear ($z \rightarrow 1$)  limit of the five-point amplitudes.
Given the universality of gauge-theory scattering amplitudes in the soft limit and the principle of leading transcendentality~\cite{Kotikov:2002ab}, we speculate that the two-loop tripole terms in the generalized splitting amplitudes in QCD are \emph{identical to} what we obtain in $\mathcal{N}=4$ super Yang-Mills theory.

Our explicit results for the space-like collinear splitting amplitudes resolves the potential issue with factorization violation in physical processes at NNNLO.
We show for the first time that the dipole terms in $\mathbf{Sp}$ evaluate to a pure phase and cancel at the level of squared amplitudes, whereas the tripole terms coincide with the prediction from soft-collinear limit. Hereby we demonstrate that factorization violating effects in the space-like collinear limit cancel at the level of color summed squared amplitudes at NNNLO in $\mathcal{N}=4$ super Yang-Mills theory, going beyond the previous considerations from infrared-pole factorization~\cite{Forshaw:2012bi} and soft gluon factorization~\cite{Dixon:2019lnw}.

\section{Conclusion and outlook}

In this \emph{Letter}, we initiated a systematic study of amplitude-level spacelike collinear factorization breaking effects. 
We obtained for the first time two-loop spacelike generalized splitting amplitudes in ${\cal N} = 4$ SYM. As a close cousin of QCD, this theory shares many similarities in the infrared limit, and therefore provides useful reference to future study for actual QCD processes. Furthermore, based on the leading transcendentality principle, we expect the results 
 obtained here also predict the leading transcendental part of the generalized spacelike splitting amplitude in QCD.

We showed that while factorization violation does exist at the amplitude level, those terms cancel at the unpolarized squared amplitude for NNNLO corrections. On the other hand, for polarized observable such as the single-spin asymmetry considered in \cite{Collins:2007nk}, our results provide concrete input for studying its factorization violation at two loops.

Our results demonstrate for the first time that factorization violation can be studied using the remarkable data of high-multiplicity scattering amplitudes. In order to do so, 
we developed a systematic methods for both taking the spacelike collinear limit of pentagon functions, as well as for taking discontinuity from integrated amplitudes. 
These methods will be valuable when considering other processes. For example, they may be used to test our conjecture about the spacelike QCD splitting amplitudes.

\section{Acknowledgment}

We thank Dmitri Chicherin, Iain Stewart and Simone Zoia for insightful discussions. We also acknowledge S\'ergio Carr\^olo and Wen Chen for their early participant in this project. This work received
funding from the European Research Council (ERC) under the European Union’s Horizon
2020 research and innovation programme (grant agreement No 725110), {\it Novel structures
in scattering amplitudes}, and under the European Union’s Horizon Europe research and
innovation programme (grant agreement No 101040760), {\it High-precision multi-leg Higgs and
top physics with finite fields} (FFHiggsTop).  KY is supported by the National Natural Science Foundation of China under Grant No. 12357077 and would also like to thank the sponsorship from Yangyang Development Fund. YZ is supported from the NSF of China through
Grant No. 12047502, 12247103, and 12075234. HXZ is supported from the NSF of China through Grant No. 11975200, Startup Grant from Peking University, and Asian Young Scientist Fellowship.

\appendix

\bibliographystyle{h-physrev}
\bibliography{haupt_kurz.bib}

\end{document}